\PassOptionsToPackage{dvipsnames,svgnames,table}{xcolor}
\documentclass[final,authoryear]{elsarticle}

\usepackage[utf8]{inputenc} 
\usepackage[T1]{fontenc}    
\usepackage{booktabs, threeparttable, multirow, siunitx}   
\usepackage{amsfonts}       
\usepackage{nicefrac}       
\usepackage{microtype}      
\usepackage{lipsum}
\usepackage[final]{changes}   
\definechangesauthor[name=Vygintas Gontis, color=red]{clc} 

\usepackage{graphicx}   
\usepackage{makecell}   
\usepackage{amsmath,amssymb,array,bm}
\pdfoutput=1
\usepackage{tabularx}
\usepackage{changepage}

\usepackage{textcomp,marvosym}
\usepackage[right]{lineno}
\usepackage{microtype}
\DisableLigatures[f]{encoding = *, family = * }
\usepackage[table]{xcolor}
\usepackage{array}
\usepackage{siunitx}

\usepackage[hidelinks]{hyperref}
\journal{International Review of Economics and Finance}

\newcolumntype{+}{!{\vrule width 2pt}}

\newlength\savedwidth

\makeatletter

\begin{document}

\begin{frontmatter}
\title{Panel regression for the GDP of the Central and Eastern European countries using time-varying coefficients\tnoteref{t1}}
\tnotetext[t1]{This is the accepted manuscript of an article published in \emph{International Review of Economics and Finance} 110 (2026) 105566. DOI: \url{https://doi.org/10.1016/j.iref.2026.105566}.}

\author[aff2,aff3]{Vygintas Gontis\corref{cor1}}
\cortext[cor1]{Corresponding author}
\ead{vygintas@gontis.eu}
\author[aff1]{Lesya Kolinets}

\address[aff1]{Vilnius Gediminas Technical University (VILNIUS TECH), Lithuania. ORCID: \href{https://orcid.org/0000-0002-7005-0519}{0000-0002-7005-0519}. \texttt{lesya.kolinets@vilniustech.lt}}
\address[aff2]{Institute of Lithuanian Scientific Society, J. Basanavi\v{c}iaus 6, Vilnius, Lithuania. ORCID: \href{https://orcid.org/0000-0002-1859-1318}{0000-0002-1859-1318}. \texttt{vygintas@gontis.eu}}
\address[aff3]{Institute of Theoretical Physics and Astronomy, Vilnius University, Saul\.{e}tekio al. 3, 10257 Vilnius, Lithuania.}

\begin{abstract}
The integration of Central and Eastern European (CEE) countries into the European Economic Area serves as a valuable experiment for the regional economic development theory. The long-lasting convergence of these economies with more advanced Western Europe exhibits a few standard features and varying policies implemented. Even the Baltic countries, which started from very similar starting positions, demonstrate their unique trajectories of development.
We propose a panel data regression model that allows coefficients to vary over time, offering a method to compare the contributions of several macroeconomic factors to the GDP growth of CEE countries. In particular, we regress the annual change of GDP per capita in PPP terms as a function of achieved GDP, price, trade, investment, and debt levels. Although other authors have extensively investigated selected regressors, the quantitative comparison of their contribution to GDP growth is a new result of the proposed method. Time-varying slope coefficients in this approach describe the external economic environment in which countries implement their own policies. The panel consists of 11 Central and Eastern European countries (Bulgaria, Czechia, Estonia, Croatia, Latvia, Lithuania, Hungary, Poland, Romania, Slovenia, and Slovakia), which have been observed annually from 1995 to 2024. While the main selected factors of this investigation contribute to economic growth, in agreement with previous findings, the role of private debt appears vital in determining growth pace. Our results serve as an argument for the further theoretical and empirical investigation of private debt contribution to economic growth.
\end{abstract}

\begin{keyword}

Economic growth \sep Central and Eastern Europe \sep Baltic countries  \sep Panel regression using time-varying coefficients \sep Private debt
\end{keyword}

\end{frontmatter}

\section{Introduction}

\added[id=clc]{Central and Eastern European (CEE) economies provide one of the clearest recent examples of economic convergence within an integrated regional market. Since the beginning of market transition and subsequent accession to the European Union (EU), these countries have experienced substantial institutional transformation, deeper trade and financial integration, and rapid income catch-up relative to Western Europe. At the same time, their growth trajectories have remained heterogeneous, indicating that convergence in the region cannot be explained by common accession effects alone \citep{Holobiuc2020EFAJ,Alemu2024AO}.} 
\added[id=clc]{This heterogeneity is especially important in a region shaped by repeated structural shocks. Over the period 1995-2024, CEE economies underwent pre-accession restructuring, EU accession, the global financial crisis, the post-crisis recovery, the COVID-19 shock, and the post-pandemic inflationary adjustment. These episodes likely altered the strength and even the direction of the relationship between macroeconomic fundamentals and growth. In such a setting, the assumption of constant coefficients may be too restrictive. Time-varying coefficient models provide a more flexible framework by allowing the effects of explanatory variables to evolve with the macroeconomic environment, rather than imposing a single average relationship across the entire sample period \citep{Hastie1993JRSSSBSM,Li2011EJ,Atak2023ET,Wang2025JE}. }
\added[id=clc]{A particularly informative case within the region is the Baltic sub-group. Estonia, Latvia, and Lithuania started the post-socialist transition from broadly similar political and institutional conditions, yet their growth paths became increasingly differentiated after the 2008 crisis. This pattern suggests that differences in the macro-financial structure of growth may matter even among countries with closely related historical trajectories. In this context, private debt deserves special attention, as a growing body of research shows that private debt booms can generate financial fragility, debt overhang, and persistent costs for the real economy \citep{Verner2019SSRN}.
Private debt should also be considered in the context of the broader finance-growth nexus. \citet{DUCTOR2015393} show that the growth effect of financial development depends on the expansion of private credit relative to real-sector output, implying that financial deepening may become less beneficial when it is not matched by productive activity. In the European context, \citet{BAHADIR2017101} document convergence in household and business credit, particularly among transition economies, but show that this process was driven mainly by household credit, which may limit its growth benefits. \citet{CANER2021694} further demonstrate that public and private debt affect growth in an endogenous and nonlinear manner. These studies support the view that private debt is not merely a passive control variable, but a potentially time-varying source of macroeconomic vulnerability in catching-up economies.} 
\added[id=clc]{The present paper addresses a clear empirical gap. We decompose GDP per capita growth across all 11 EU-acceding CEE countries over the full 1995-2024 period and across seven macroeconomic factors. Existing studies on CEE growth have typically relied on static panel models, shorter time horizons, or narrower country samples, which makes it difficult to capture how the role of key macroeconomic factors changes across major structural episodes. }
\added[id=clc]{The objective of this paper is to identify and compare the time-varying contributions of seven macroeconomic factors to GDP per capita growth in 11 CEE economies: Bulgaria, Czechia, Estonia, Croatia, Latvia, Lithuania, Hungary, Poland, Romania, Slovenia, and Slovakia. Using a linear regression with varying coefficients (LRVC) approach, the paper relates annual changes in GDP per capita in purchasing power parity terms to achieved GDP level, price level, foreign direct investment, trade openness, gross capital formation, central government debt, and private debt.}
\added[id=clc]{The paper is guided by the expectation that macroeconomic determinants are time-varying and that private debt may be an important source of post-2008 growth divergence in CEE. This expectation is consistent with the argument that while credit expansion may support development at moderate levels, excessive leverage can become a drag on growth once it increases financial vulnerability and post-crisis adjustment costs \citep{Verner2019SSRN}.} 
\added[id=clc]{The novelty of the paper lies in four main aspects. First, unlike prior empirical studies on CEE growth that mainly rely on conventional panel, cointegration, ARDL, or convergence frameworks \citep{Gherghina2019Sust,Osinska2025-nm}, this paper applies a linear regression with varying coefficients framework to analyze GDP per capita growth in CEE countries. Second, to the best of our knowledge, it is the first study to examine the full set of 11 EU-acceding CEE economies within a single time-varying coefficient panel framework; previous studies either analyze the same CEE-11 group over a shorter period or focus on narrower subregional samples, such as the Baltic states \citep{Dritsaki2020AEFR}. Third, the paper covers a long time horizon, 1995-2024, thereby encompassing pre-accession restructuring, EU accession, the global financial crisis, the COVID-19 shock, and the post-pandemic period, whereas earlier studies usually rely on substantially shorter windows, such as 1994-2019 \citep{Alemu2024AO} or 2003-2016 \citep{Gherghina2019Sust}. Fourth, the paper offers a new interpretation of post-2008 Baltic divergence by highlighting the private debt channel as a potential source of cross-country differentiation, extending the broader literature that identifies private debt booms as a source of macroeconomic fragility and weaker real-side performance \citep{Mian2017-lg,Dinh2019JRFM,Park2021-on,Gu2022EM}.}
\added[id=clc]{The remainder of the paper is organized as follows. Section 2 reviews the literature. Section 3 presents the econometric framework. Section 4 describes the data and variables. Section 5 reports the empirical results. The final section concludes.}

\section{\added[id=clc]{Literature review} \label{sec:review}}
\subsection{\added[id=clc]{Economic convergence and heterogeneity in Central and Eastern Europe} \label{sec:EconomConverg}} 
\added[id=clc]{The literature on Central and Eastern Europe increasingly treats the region as a group of heterogeneous catching-up economies rather than as a uniform regional category. Empirical evidence confirms that CEE countries narrowed part of their income gap relative to Western Europe. However, the pace, durability, and structural composition of convergence differed markedly across countries and over time \citep{Holobiuc2020EFAJ,Alemu2024AO}. Related research shows that some Eastern European economies continued to catch up while others became partially stuck, implying that convergence in the region is conditional rather than automatic \citep{Cieslik2020-bs,Konya2023-he}. Comparative analyses likewise indicate that productivity, capital accumulation, policy quality, and institutional settings contributed unevenly to growth across CEE economies, particularly around and after the global financial crisis \citep{Jesic2023-xz,Shkolnykova2025-op}.
This heterogeneity is not limited to differences in income levels or growth rates, but also reflects broader differences in institutional conditions and development paths. Recent evidence indicates that economic freedom, GDP per capita, and human development are linked in the long run, while the region itself is better understood as a set of distinct institutional clusters rather than as a homogeneous bloc \citep{Osinska2025-nm}. Accordingly, differences in growth performance across CEE countries cannot be reduced to convergence effects alone. They also reflect variation in policy environments, institutional quality, and the broader configuration of development models operating within the region.}
\added[id=clc]{Nevertheless, variation in the speed of convergence across CEE countries remains insufficiently explained by convergence theory alone. The problem has motivated a broader literature on the macroeconomic factors shaping GDP growth, including price dynamics, foreign direct investment (FDI), trade openness, capital formation, and both public and private debt. Much of this literature relies on cross-country or panel-data frameworks to identify the main correlates of growth in developed and developing economies. The key question, therefore, is not only whether CEE countries have converged, but also which macroeconomic and financial forces have shaped the differences in their growth trajectories.}
\subsection{\added[id=clc]{Macroeconomic determinants of growth in catching-up economies} \label{sec:MacroekonomikDet}} 
\added[id=clc]{A structural rationale for including the price level in growth regressions for catching-up economies is provided by the Penn-Balassa-Samuelson mechanism \citep{Balassa1964JPE}. As productivity rises more rapidly in tradable sectors, wages tend to increase economy-wide, which in turn raises the relative prices of non-tradables and the aggregate price level. Earlier studies for Central and Eastern Europe document the relevance of this channel for real convergence. However, they also show that productivity differentials explain only part of observed inflation differentials across countries \citep{Lojschova2003IAS,Mihaljek2003BISWP,Armendariz2025IMF}. Incorporating price levels into growth analysis, therefore, helps distinguish productivity-consistent real appreciation from forms of price adjustment that may weaken external competitiveness \citep{Hassan2016JIE}.}
\added[id=clc]{The broader inflation literature also indicates that macroeconomic stability remains important for sustainable growth. \citet{Barro2013AEF} treats inflation as a key conditioning variable in long-run growth analysis. At the same time, \citet{He2023EM} finds that the relationship between inflation and growth is nonlinear, resembling an inverted U, with the marginal effect of inflation on growth turning negative at relatively low levels. For EU economies, recent evidence indicates that the direct adverse effect of inflation on growth is not always robust, but that tighter monetary conditions associated with inflation control can weigh on economic performance \citep{Pappas2025Economies}. \citet{Batrancea2021-iv} likewise shows, using panel evidence for EU countries, that GDP is positively associated with economic sentiment. In contrast, inflation exerts a negative effect, highlighting the macroeconomic relevance of instability and sentiment dynamics during the COVID-19 period. Various arguments imply that price developments matter not only through convergence-related real appreciation, but also through their interaction with monetary conditions and macroeconomic stability.}
\added[id=clc]{Some studies explicitly examine price-level convergence alongside income convergence. They show that poorer European economies often start with lower price levels and experience faster price increases as they catch up with more advanced countries \citep{Rogers2007JME}. More recent Baltic-focused evidence likewise suggests that real price appreciation is not necessarily harmful, but may become problematic when it outpaces productivity growth and erodes competitiveness \citep{Armendariz2025IMF}. From this perspective, sustained long-run growth in open economies depends not simply on higher price levels but on maintaining consistency between price adjustment, productivity growth, and external competitiveness.}

\added[id=clc]{Foreign direct investment is widely discussed in the growth literature as a channel through which host economies acquire capital, technology, managerial know-how, and access to international production networks. Recent empirical evidence supports a positive FDI-growth nexus, particularly when productivity gains accompany FDI. Using an unbalanced panel of 90 middle-income countries over 1990–2020, \citet{Le2024HSSC} find that a percentage increase in foreign direct investment is associated with higher economic growth and that total factor productivity further strengthens this relationship. This result is consistent with the view that FDI contributes to growth not only through capital inflows, but also through technological spillovers and improvements in production efficiency.}
\added[id=clc]{In the context of Central and Eastern Europe, the literature likewise links FDI to restructuring, productivity upgrading, and the broader catching-up process. \citet{Acaravci2012JEF}, analyzing ten new EU member states, document long-run relationships among FDI, exports, and economic growth, thereby supporting the view that external integration is an important component of growth performance in transition economies. In a broader discussion of the region, \citet{Popescu2014Sust} also emphasizes the role of FDI in productivity convergence, economic restructuring, and FDI-assisted development patterns across CEE countries. At the same time, the empirical relationship is not necessarily linear. Using panel data for 11 CEE countries over 2003–2016, \citet{Gherghina2019Sust} find evidence of a nonlinear association between FDI inflows and GDP per capita, suggesting diminishing returns at very high levels of FDI. Taken together, these studies indicate that FDI is an important correlate of growth in CEE economies, but that its effect depends on host-country conditions and may vary with the scale and quality of inflows.}
\added[id=clc]{Trade openness is widely discussed as a channel through which economies benefit from specialization, competitive pressures, and the diffusion of knowledge through international exchange. In the Baltic context, \citet{Dritsaki2020AEFR} argues that greater openness is expected to stimulate economic activity because trade facilitates the transmission of know-how, innovation, and productivity improvements. Using a multimodal panel framework for 36 European countries, \citet{Batrancea2022-wj} further shows that exports and imports are closely associated with alternative measures of economic growth. In a post-pandemic setting, \citet{Spahiu2022JLIA} likewise finds that exports had a statistically significant positive effect on GDP growth in Kosovo, while imports displayed a slight inverse relationship. Taken together, these studies suggest that openness-related variables are relevant for growth performance, although their effects may vary across countries and institutional settings. \citet{Rodrik2018JIBP} provides a broader globalization perspective, highlighting that deeper integration reshapes domestic adjustment processes and therefore may generate both opportunities and constraints for catching-up economies.}

\added[id=clc]{Gross capital formation (GCF) is commonly used as a proxy for domestic investment in fixed assets and productive capacity, and its relevance is consistent with both neoclassical and endogenous growth frameworks, in which capital deepening raises productivity and potential output. Empirical evidence generally treats capital formation as an important component of the growth process. However, the magnitude and even the direction of its effect may vary across countries and institutional settings \citep{Onyinye2017AJEBA,Aslan2021ESPRI}. In the OECD context, \citet{Morina2023EJGE} find a positive relationship between domestic investment and economic growth and report a long-term causal relationship between GDP and gross fixed capital formation. Broader cross-country evidence also indicates that the growth effects of capital accumulation are heterogeneous. Using panel data for 124 countries, \citet{Topcu2020RP} show that gross capital formation has a positive impact on growth in high-income economies, while the broader relationship differs across income groups. At the country level, the evidence is more mixed: \citet{Onyinye2017AJEBA}, analyzing Nigeria, identifies a stable long-run relationship between capital formation and growth, but does not find a statistically significant effect of gross capital formation on output. Taken together, these studies suggest that capital formation remains a theoretically central and empirically relevant growth determinant, but that its contribution depends on country characteristics, development level, and the broader macroeconomic environment.}

\added[id=clc]{The debt-growth relationship is generally regarded as complex and potentially nonlinear. In the case of public debt, moderate borrowing can support output when it finances productive expenditure, such as infrastructure, education, and other growth-enhancing investments. However, a substantial empirical literature suggests that excessive public debt is more often associated with weaker medium-term economic performance, especially when debt accumulation becomes persistent or fiscal space is constrained \citep{Mudayen2025Economies,Asteriou2021JEF,Musa2023Mathematics}. Earlier influential studies also report adverse growth effects of high public debt, although the existence and precise location of threshold effects remain debated across samples and estimation strategies \citep{Reinhart2010RePec,Checherita-Westphal2012EER}. More recent evidence points in the same general direction. Using panel quantile methods for 127 developing countries, \citet{Mudayen2025Economies} show that public debt is associated with weaker growth, particularly at higher debt levels. In a broader historical setting, \citet{Jalles2024-tv} further demonstrates that both public and private debt surges are typically followed by persistently weaker economic growth. Accordingly, including government debt in growth models helps capture the role of fiscal imbalances and debt-related constraints in shaping macroeconomic performance.}
\added[id=clc]{Private debt, comprising the liabilities of households and non-financial corporations, has become even more central to contemporary macro-financial research on growth. The experience of the global financial crisis, subsequent credit expansions in emerging markets, and later shocks, including the COVID-19 period, intensified interest in the ways private leverage interacts with real economic performance. A growing body of evidence suggests that credit expansion may support activity in the short run, but can become growth-reducing when leverage rises beyond sustainable levels or when debt-financed booms amplify subsequent adjustment costs. \citet{Verner2019SSRN} shows that private debt booms often generate higher long-run costs than benefits through financial fragility and resource misallocation. \citet{Park2021-on} similarly find that household and corporate debt accumulation can impair real economic outcomes, especially when vulnerabilities intensify. \citet{Gu2022EM} further demonstrates that the debt-growth relationship differs across public and private debt categories and identifies an inverted-U pattern for private debt, with adverse effects emerging once leverage exceeds threshold levels. Their estimates place the turning point for total private debt at roughly 125\%–150\% of GDP. Evidence from developing economies also indicates that the private credit-growth nexus is dynamic rather than uniform: \citet{Dinh2019JRFM} report that private-sector credit may hinder growth in the short run while supporting it over longer horizons. Related work by \citet{Mian2017-lg} reinforces this interpretation by showing that credit supply expansions can stimulate demand temporarily but also deepen subsequent downturns. Taken together, these findings suggest that private debt should not be treated as a passive financial control. Rather, it is a distinct and potentially time-varying source of macroeconomic vulnerability whose growth effects depend on leverage levels, debt composition, and crisis exposure.}
\added[id=clc]{A broader structural perspective also suggests that long-run growth should increasingly be understood in the context of changing development models. In this respect, \citet{Batrancea2021-iv} argues that economic growth must be considered alongside the transition toward a green economy, where sustainable development, green investment, and supportive policy frameworks become increasingly important for future economic performance. Although this perspective is broader than the macro-financial focus of the present study, it reinforces the general point that growth dynamics are shaped not only by traditional macroeconomic variables but also by wider structural transformations.}
\subsection{\added[id=clc]{Evolving effects of macroeconomic determinants on growth} \label{sec:EvolvingEff}} 
\added[id=clc]{The literature reviewed above identifies the main macroeconomic and financial determinants of economic growth. However, it also indicates that the strength and direction of these effects may change across successive structural episodes. The CEE region has undergone successive phases of transition, EU accession, crisis adjustment, post-crisis recovery, and post-pandemic rebalancing, which are likely to have altered both the strength and direction of the effects of key growth determinants. In such a setting, empirical frameworks that impose constant average coefficients across the full sample may obscure the variation most relevant to explaining heterogeneous growth outcomes across CEE economies. This concern is consistent with the broader growth-econometrics literature, which increasingly emphasizes that parameter heterogeneity should be modeled explicitly rather than absorbed into static average effects (\citep{Feng2022-iv}.
The econometric foundation for doing so originates in the literature on varying coefficients. \citet{Hastie1993JRSSSBSM} introduced varying-coefficient models as a flexible framework in which regression parameters are allowed to evolve systematically with an index variable. \citet{Fan1999AS} subsequently provided a broader statistical treatment of estimation and inference in such models, establishing varying-coefficient specifications as a rigorous alternative to fixed-parameter regressions when structural relationships are expected to shift over time. In panel-data applications, \citet{Li2011EJ} extend this logic by proposing nonparametric time-varying coefficient models with fixed effects, in which slope parameters are estimated through kernel-weighted procedures after controlling for unobserved heterogeneity. Their framework is particularly relevant for macroeconomic panels, where the effects of explanatory variables may evolve gradually rather than change abruptly.}
\added[id=clc]{More recent work has broadened this class of models to reflect the complexity of macro panel data better. \citet{Wang2025JE} develop a panel framework with time-varying interactive fixed effects, allowing both regression coefficients and latent factor loadings to change smoothly over time. The model is especially useful in macroeconomic settings where common shocks, financial conditions, and policy environments evolve continuously rather than abruptly. \citet{Dong2021JBES} likewise extend the varying-coefficient panel model to settings characterized by nonstationarity and partially observed factor structures. In contrast, \citet{Chang2025-gd} incorporates time-varying coefficients into spatial panel settings, allowing for the joint analysis of slope instability and spillover effects. These contributions reinforce the view that time-varying parameterization is not simply a technical refinement, but a substantively important way of modeling economic relationships in panels shaped by changing common conditions and heterogeneous transmission mechanisms.}
\added[id=clc]{The recent econometric literature has also strengthened the case for treating coefficient stability as an empirical issue rather than as an assumption. \citet{Atak2023ET} develop nonparametric specification tests for homogeneous and stable coefficient structures in panel data models, providing a direct statistical framework for assessing whether time variation is warranted. The importance of methodological choice is likewise evident in the convergence literature itself. \citet{Desli2020EM} show that empirical conclusions about convergence are sensitive to the estimation strategy employed. In contrast, \citet{Li2016-ev} demonstrates that allowing coefficients to vary can alter the interpretation of cross-country output convergence. Taken together, these studies suggest that the empirical analysis of growth in CEE economies should be based on a framework capable of capturing evolving rather than fixed macroeconomic relationships.}
\added[id=clc]{Against this background, a time-varying coefficient panel framework is particularly well suited to the present study. If the effects of key macroeconomic determinants differ across the phases of convergence, crisis, and adjustment, then a static specification may yield only an average approximation that masks the mechanisms driving divergence within the region. This consideration is especially important in the CEE context, where the literature already points to heterogeneous post-crisis outcomes and to the growing macro-financial relevance of debt, particularly private debt, in shaping medium-term performance. It also provides the immediate analytical foundation for the study’s formal hypotheses.}
\subsection{\added[id=clc]{Research hypotheses} \label{sec:Hypotheses}} 
\added[id=clc]{Based on the reviewed literature and the identified empirical gap, the study formulates the following hypotheses. H1. The effects of the selected macroeconomic determinants on GDP per capita growth in CEE countries vary over time and particularly concentrate around the identified structural episodes (2008, COVID). H2. Private debt is expected to be a strong adverse correlate of GDP per capita growth in CEE countries, particularly after 2008. H3. Differences in private debt dynamics help explain the post-2008 divergence in growth trajectories among the Baltic countries.}
\section{The time-varying coefficient panel model \label{sec:regression}}
We employ a panel data regression model that allows coefficients to vary over time. In particular, we model GDP per capita (in purchasing power parity, or PPP, terms) as a function of several economic factors with time-varying slope coefficients. The panel consists of 11 Central and Eastern European countries (Bulgaria, Czechia, Estonia, Croatia, Latvia, Lithuania, Hungary, Poland, Romania, Slovenia, and Slovakia), which have been observed annually from 1995 to 2024.

A panel model with variable coefficients allows us to capture structural changes and the evolution of relationships over the period 1996–2024 \citep{Atak2023ET}. Such an approach is essential for the transition economies of the CEE region, which have undergone significant transformations, including accession to the European Union in the 2000s and the 2008 global financial crisis. Transformation events may have altered the impact of key factors on economic growth, making models with time-dependent parameters particularly relevant.

For the comparison of GDP dynamics across CEE countries, we use a panel of time-varying linear regression coefficients. 

\begin{equation}
\Delta y_{i,t}=\alpha_t + \sum_{j=1}^d \beta_{t-1,j} x_{i,t-1,j} +e_{i,t},\qquad i=1,...,N,\quad t=1,...,T
\label{Eq:regression}
\end{equation}
where $\Delta y_{i,t}$ represents GDP/pc annual change in country $i$, at year $t$, $\bm{x}_{i,t-1}$ is the vector of $d$ regressed variables, and the vector $\bm{\beta}_t=\{\beta_{t,1},...,\beta_{t,d}\}$ represents linear time-varying regression coefficients. The error process we denote as $\{e_{i,t}\}$. In this research, we will use seven regressed variables: X1 - GDP per capita in PPP, X2 - Price level index, X3 - Accumulated foreign direct investment, X4 - Trade, X5 - Accumulated gross capital formation, X6 - Central Government debt, and X7 - Private debt. The data description follows. This data we use to get matrix $\bm{x}$, for the eleven countries, 29 time steps $\{1995,....,2023\}$, and seven, $d=7$, economic growth financial components $Xj$. We use the same GDP per capita data in PPP to calculate the matrix $\bm{\Delta y}$ for the 29 time intervals from 1995 to 2024 and the selected 11 countries.

We can estimate the time-varying regression coefficients $\beta_{t-1,j}$ and the vector $\bm{\alpha}_t$ common across countries using panel least squares. These empirically defined regression parameters can be used to evaluate the regressed $\bm{\Delta y}$, error terms $\{e_{i,t}\}$, and the contribution of all components $Xj$ to the $\Delta y$ defined as $\Delta y_{i,t,j}$
\begin{equation}
\Delta y_{i,t,j}=\frac{\alpha_t}{d}+\beta_{t-1,j} x_{i,t-1,j}, \qquad \Delta y_{i,t}=\sum_{j=1}^d \Delta y_{i,t,j}.
\label{Eq:components}
\end{equation}
For a country $i$ and  component $j$, we will plot the GDP component dynamics 
\begin{equation}
\kappa(t)=\kappa_{i,t,j}=GDP_{1995}/d+\sum_{\tau=1996}^t \Delta y_{i,t,j},
\label{Eq:components2}
\end{equation}
and the relative contribution of private debt to the GDP is defined as
\begin{equation}
\lambda(t)=sgn(\kappa_{i,t,7}) \kappa_{i,t,7}^2/\sum_{j=1}^d \kappa_{i,t,j}^2.
\label{Eq:relcontr7}
\end{equation}
We use three different estimates of $\Delta y_{i,t}$ seeking to evaluate the accuracy of the modeling. Empirical one $\Delta y_{i,t}^e=x_{i,t,1}-x_{i,t-1,1}$, regressed one $\Delta y_{i,t}^r=\Delta y_{i,t}$ calculated from the Eq. \eqref{Eq:regression}, and full regressed $\Delta y_{i,t}^{fr}$, when we replace in calculation all empirical values of $x_{i,t-1,1}$ by the accumulated values of $\Delta y_{i,t}$ obtained in previous time steps of estimation. It is evident that the error term in Eq. \eqref{Eq:regression} can be expressed as $e_{i,t}=\Delta y_{i,t}^r-\Delta y_{i,t}^e$. The results of these three estimates we will illustrate as  
\begin{align}
GDP Data(t) &=GDP_{1995}+\sum_{\tau=1996}^t\Delta y_{i,\tau}^e, \label{Eq:GDPData}\\
GDP Regr.(t) &=GDP_{1995}+\sum_{\tau=1996}^t\Delta y_{i,\tau}^r, \label{Eq:GDPRegr}\\
GDP Regr.Full(t) &=GDP_{1995}+\sum_{\tau=1996}^t\Delta y_{i,\tau}^{fr}, \label{Eq:GDPRefrFull}\\
e_{i,t}^a =\sum_{\tau=1996}^t e_{i,\tau} &=GDP Regr.(t)-GDP Data(t)\label{Eq:errorAccumulated}
\end{align}

\subsection{\added[id=clc]{Orthogonalization of regressors} \label{sec:orthog}}
\added[id=clc]{The seven regressors $X_{1},\dots,X_{d}$ ($d=7$), which we investigate in this contribution, are levels of macroeconomic
factors that share a common upward trend in the panel, so their pooled
correlations are high. The resulting collinearity inflates the variance of
the per-year slope estimates $\hat{\beta}_{t-1,j}$ in equation~(1). We
address it by applying a single common linear transformation to the
regressors, fitting the LRVC equation in the resulting orthogonal basis,
and translating the estimated coefficients back into the original
variables. The transformation is invertible, and we retain all $d$
components, so the procedure does not change the fit -- only its numerical
conditioning -- and the final results are reported in the units of
$X_{1},\dots,X_{7}$.}

\paragraph{Linear transformation}
\added[id=clc]{With the pooled mean and standard deviation}
\begin{equation}
\mu_{j} \;=\; \frac{1}{NT}\sum_{i,t} x_{i,t,j},
\qquad
\sigma_{j}^{2} \;=\; \frac{1}{NT-1}\sum_{i,t}\bigl(x_{i,t,j}-\mu_{j}\bigr)^{2},
\label{eq:mu_sigma}
\end{equation}
define standardized observations
$z_{i,t,j}=(x_{i,t,j}-\mu_{j})/\sigma_{j}$. Let $\Omega\in\mathbb{R}^{d\times d}$
\added[id=clc]{be the pooled correlation matrix of $z$, with eigendecomposition}
\begin{equation}
\Omega \;=\; W\Lambda W^{\top},
\qquad
W^{\top}W = I_{d},
\qquad
\Lambda = \operatorname{diag}(\lambda_{1},\dots,\lambda_{d}).
\label{eq:eig}
\end{equation}
\added[id=clc]{The orthogonal regressors $P_{1},\dots,P_{d}$ are obtained by applying the
same loading matrix $W$ to every observation:}
\begin{equation}
p_{i,t,k} \;=\; \sum_{j=1}^{d} W_{j,k}\, z_{i,t,j},
\qquad k=1,\dots,d.
\label{eq:Pdef}
\end{equation}
By construction
$\frac{1}{NT-1}\sum_{i,t} p_{i,t,k}\,p_{i,t,\ell} = \lambda_{k}\,\delta_{k\ell}$,
so the $P_{k}$ are mutually uncorrelated in the pooled sample.

\paragraph{LRVC in the orthogonal basis}
\added[id=clc]{Substituting $p_{i,t-1,k}$ for $x_{i,t-1,j}$ in equation~(1) gives}
\begin{equation}
\Delta y_{i,t} \;=\; \alpha_{t}^{P} \;+\; \sum_{k=1}^{d}
\gamma_{t-1,k}\, p_{i,t-1,k} \;+\; e_{i,t},
\label{eq:LRVC_P}
\end{equation}
\added[id=clc]{estimated year by year on the $N$-country cross-section. The design matrix
of \eqref{eq:LRVC_P} is well-conditioned because the regressors are
orthogonal.}

\paragraph{Recovery in the original variables}
\added[id=clc]{Inserting \eqref{eq:Pdef} and the definition of $z_{i,t,j}$ into
\eqref{eq:LRVC_P} and collecting terms in $x_{i,t-1,j}$ reproduces equation~(1)
with the slopes and intercept}
\begin{align}
\beta_{t-1,j} &\;=\; \frac{1}{\sigma_{j}}\sum_{k=1}^{d} W_{j,k}\,\gamma_{t-1,k},
\label{eq:beta_recover}\\
\alpha_{t}    &\;=\; \alpha_{t}^{P} \;-\; \sum_{j=1}^{d}\mu_{j}\,\beta_{t-1,j},
\label{eq:alpha_recover}
\end{align}
\added[id=clc]{or equivalently, in matrix form,
$\boldsymbol{\beta}_{t-1} = \operatorname{diag}(\boldsymbol{\sigma})^{-1}\,W\,\boldsymbol{\gamma}_{t-1}$
and
$\alpha_{t} = \alpha_{t}^{P} - \boldsymbol{\mu}^{\top}\boldsymbol{\beta}_{t-1}$.
Because $W$ is orthogonal and all $d$ components are retained, the fitted
$\Delta y_{i,t}$ and the residuals $e_{i,t}$ in \eqref{eq:LRVC_P} coincide
with those of equation~(1) up to numerical precision; the orthogonalization
serves as a preconditioner that stabilizes the per-year estimation. The
recovered $\beta_{t-1,j}$ and $\alpha_{t}$ feed unchanged into the
component decomposition (eq.~2) and the cumulative GDP path (eq.~3).}

\section{Empirical data  \label{sec:Data}}

\begin{figure}[h]
\includegraphics[width=0.5\textwidth]{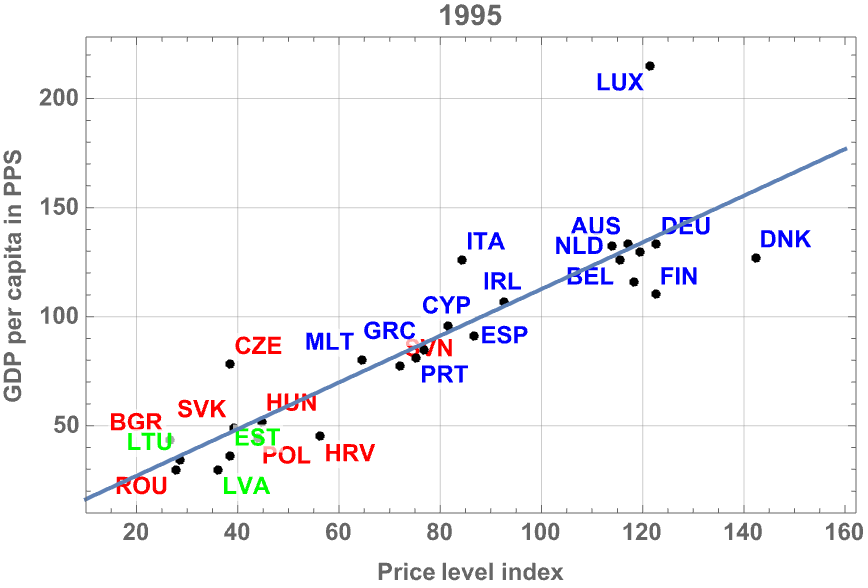}
\includegraphics[width=0.5\textwidth]{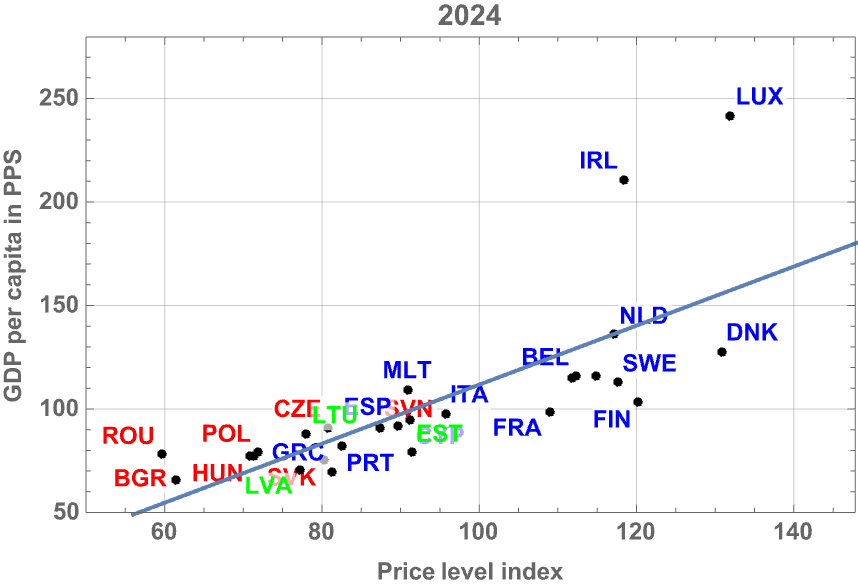}
\caption{The price index and GDP/pc PPS scatter plot for the EU countries in 1995 and 2024. Red: CEE countries except those in the Baltic region; green: Baltic countries; blue: the rest of the EU countries. The line shows the least-squares trend for all EU countries.
\label{fig1}}
\end{figure}

\begin{figure}[h]
\centering
\includegraphics[width=0.5\textwidth]{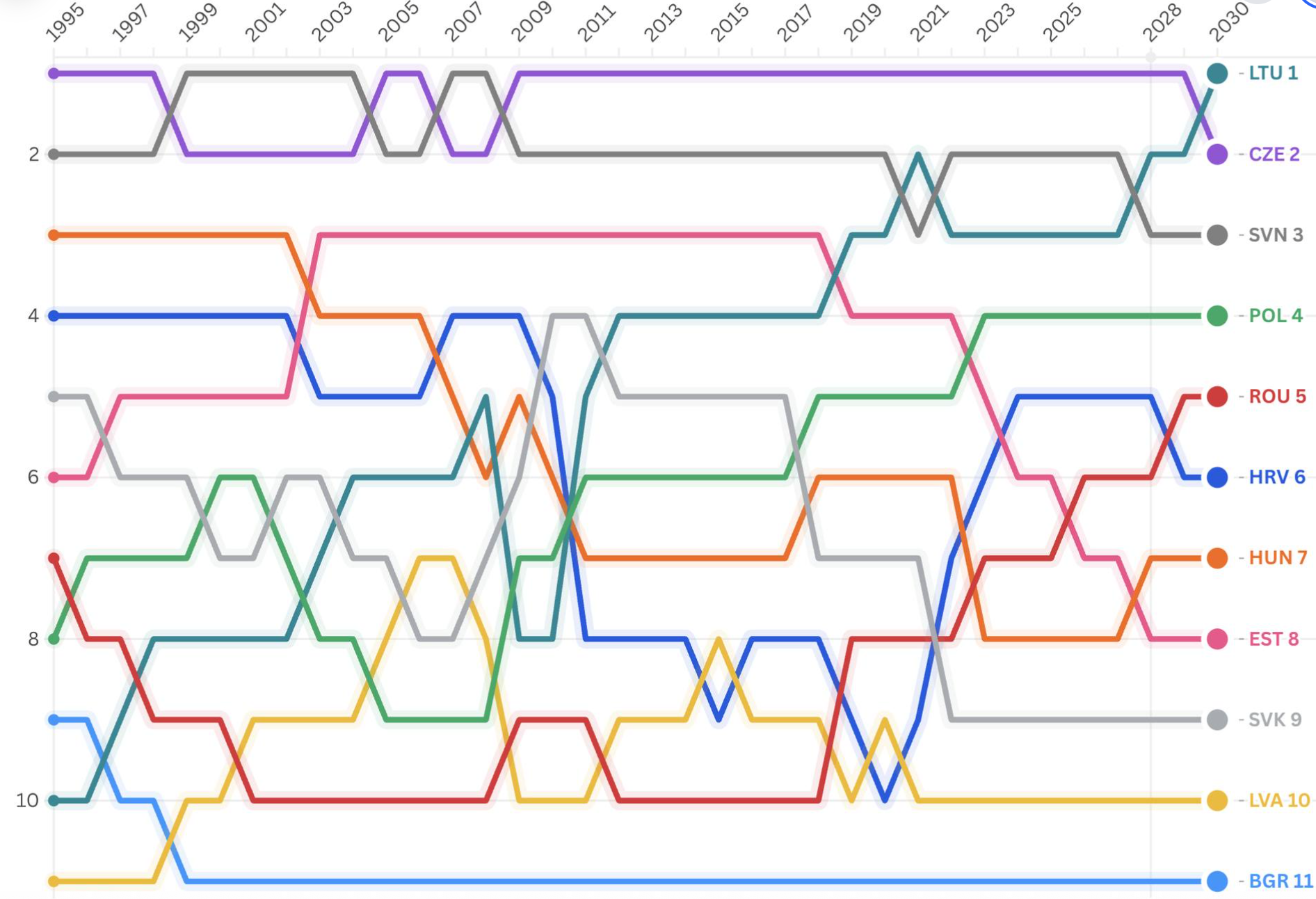}
\caption{The GDP/pc PPP rank change of the Central and Eastern European countries. 
\label{fig2}}
\end{figure}

Central and Eastern European countries experience rapid economic convergence with the rest of Europe \citep{Holobiuc2020EFAJ,Alemu2024AO}. The convergence process follows the Penn Effect, explained by the Balassa-Samuelson hypothesis. The effect is comparatively well expressed for the EU countries, as shown in Fig. \ref{fig1}.
Nevertheless, this crucial catch-up process for the EU project varies significantly across CEE countries. It is very illustrative to follow how the economic ranking of the CEE countries changes from 1995 to 2030, according to the International Monetary Fund's projected data \citep{IMF_WEO_2025_07}, as shown in Fig. \ref{fig2}. \added[id=clc]{We started the investigation in 1995, as reliable data for the CEE countries is available only from this year. Before then, CEE countries underwent significant institutional reforms to prepare for their transition to the European Economic Area. For example, the Baltic countries had to introduce their own currencies, making them convertible in international exchange.} The ranking of Lithuania changes from tenth to first, as other countries exhibit considerable variation in their rankings as well. We aim to investigate the contributions of achieved GDP, price levels, and other macroeconomic factors to economic growth, explaining the variable behavior of CEE countries relative to the Penn effect trend and their economic growth rankings.


\begin{table}[!htbp]
\centering
\caption{\added[id=clc]{Variable definitions and sources.}}
\label{tab:variables}
\small
\begin{tabularx}{\textwidth}{l X l l l}
\toprule
Symbol & Variable & Unit & Transform in (1) & Source \\
\midrule
$\Delta y$ & Annual change in GDP per capita, PPP & PPS, EU27=100 & first difference & Eurostat \\
$X_{1}$ & Volume indices of real expenditure pc & PPS, EU27=100 & level, lagged $t{-}1$ & Eurostat \\
$X_{2}$ & Price level index & EU27=100 & level, lagged $t{-}1$ & Eurostat \\
$X_{3}$ & Accumulated FDI net inflows & cum.\ \% GDP & level, lagged $t{-}1$ & World Bank \\
$X_{4}$ & Trade (exports $+$ imports) & \% GDP & level, lagged $t{-}1$ & World Bank \\
$X_{5}$ & Accumulated gross capital formation & cum.\ \% GDP & level, lagged $t{-}1$ & World Bank \\
$X_{6}$ & Central government debt & \% GDP & level, lagged $t{-}1$ & Eurostat (ESTAT) \\
$X_{7}$ & Private sector debt, consolidated & \% GDP & level, lagged $t{-}1$ & Eurostat (ESTAT) \\
\bottomrule
\end{tabularx}
\end{table}
\added[id=clc]{We organize empirical analysis relying on the macroeconomic modeling of economic convergence based on diffusion of technologies \cite{Gontis2026Arxiv}. In the neoclassical model of economic growth with two factors of productivity: labor $L$ and capital $K$ in the Cobb–Douglas production function}
\begin{equation}
Y=AL^{1-\delta}K^{\delta},
\label{eq:Growth}
\end{equation}
\added[id=clc]{$A$ denotes the total factor productivity quantifying the technology transfer from the more advanced economies. In this setting, $A$ accounts for the most significant part of economic growth in CEE countries, and the speed of convergence is quantified by a single parameter, $\gamma$, see Eq. (19) in \cite{Gontis2026Arxiv}. Do not confuse this notation with that in this article. Thus, we aim to identify a few macroeconomic factors most responsible for the pace of CEE convergence with advanced European economies. It should be evident that we have to deal with interconnected factors, such as the price level and the level of economic development, or foreign direct investment and gross capital formation. Thus, our empirical analysis must address possible multicollinearity among the selected factors.}

For the panel regression described in the section \ref{sec:regression} we use following regressed variables - economic growth factors:
X1 - Volume indices of real expenditure per capita (in PPS EU27 2020 100). This data is available at Eurostat. Purchasing power parities (PPPs), price level indices, and real expenditures for ESA 2010 aggregates. 
X2 - Price level indices (EU27 2020 100). Purchasing power parities (PPPs), price level indices, and real expenditures for ESA 2010 aggregates.
X3 - Foreign direct investment, net inflows (\% of GDP), World Bank data. The data we used were collected during the period considered. 
X4 - Trade (\% of GDP). Trade is the sum of exports and imports of goods and services. World Bank data.
X5 - Gross capital formation (\% of GDP). Gross capital formation includes acquisitions, less disposals, of produced assets for fixed capital formation, inventories, or valuables. World Bank data. The data we use is accumulated in the period considered. 
X6 - Central Government Debt, Government deficit/surplus, debt and associated data (\% of GDP), ESTAT data.
X7 - Private sector debt, consolidated - \% of GDP [tipspd20]. Consolidated non-financial corporations, households, and non-profit institutions serving households, as per ESTAT data. \added[id=clc]{See Table \ref{tab:variables} for the more concise presentation of selected variables. We provide the descriptive statistics for the empirical data used in Table \ref{tab:descrptstats}.
We evaluate the expected interdependence or regressors $X1...X7$, calculating the pairwise correlation matrix $\Omega$ of standardized observations $z_{i,t,j}$, Eq. \eqref{eq:eig}, see Table \ref{tab:corr}. The highest correlation is estimated for the pair {X1,X2} of GDP and price level. The gross capital formation, X5, shows a strong correlation with all other regressors. The observed interdependence among selected factors is an indispensable macroeconomic feature of the system investigated. Nevertheless, variance inflation factors (VIF) for X1–X7 range from 1.1 to 4.0 (mean 2.6), all below the conventional threshold of 5, indicating no harmful multicollinearity among the regressors. The regressors are moderately correlated, and the small yearly cross-sections make the per-year design poorly conditioned; we therefore apply the orthogonal preconditioner of subsection \ref{sec:orthog} via the transformation in Eq. \eqref{eq:Pdef}. Subsection \ref{sec:orthog} provides a detailed description of how we implement a single common linear transformation to the
regressors, fitting the LRVC equation in the resulting orthogonal basis,
and translating the estimated coefficients back into the original
variables. All calculations and statistical analyses were carried out in Wolfram Mathematica (14.0.0.0)}

\begin{figure}[h]
\centering
\includegraphics[width=0.4\textwidth]{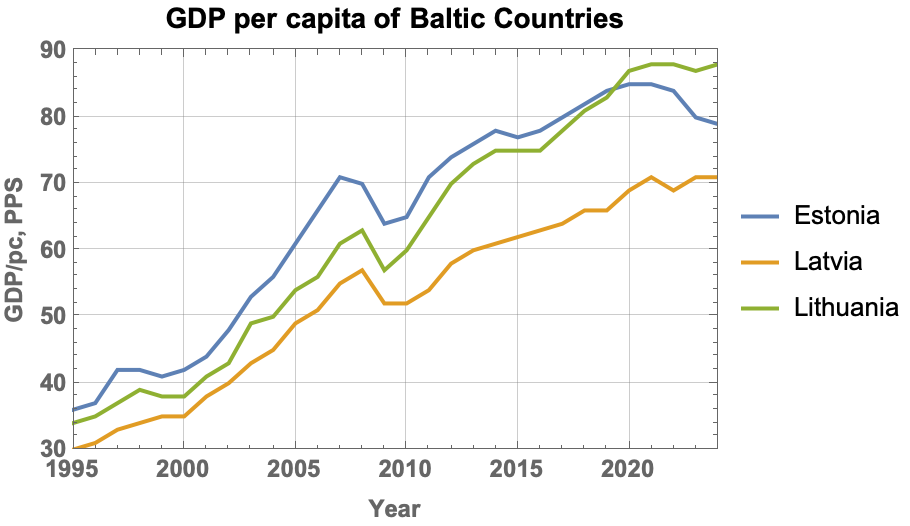}
\includegraphics[width=0.4\textwidth]{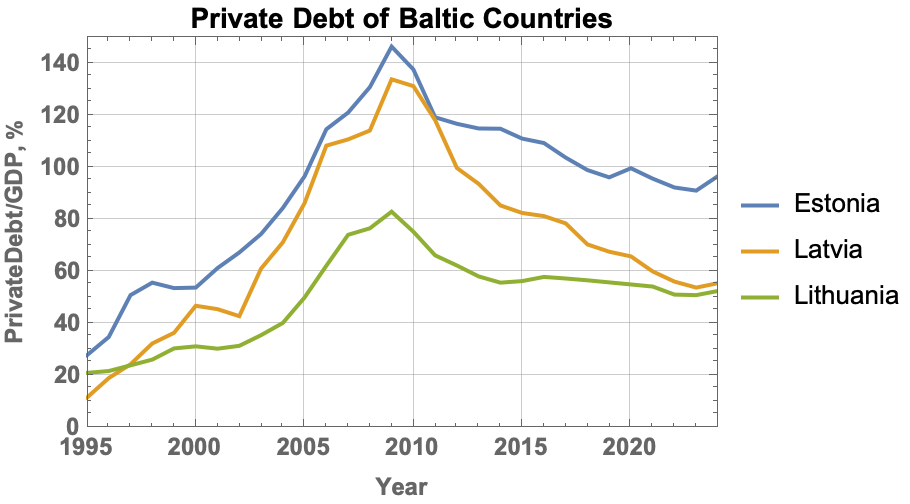}
\caption{The comparison of Baltic countries. The first sub-figure compares GDP per capita in PPS, and the second one compares private debt, \% of GDP.
\label{fig3}}
\end{figure}

In this study, we pay particular attention to the comparison of Baltic countries, as they began European integration from a very similar level of development and shared the same political heritage. The Baltic countries, which experienced a similar pace of development at the beginning of the period, exhibit some divergence in the last interval of time (after the economic crises), as seen in the first sub-figure of Fig. \ref{fig3}. We do expect to reveal macroeconomic factors responsible for the observed divergence. 

\begin{table}[!htbp]
    \centering
    \caption{\added[id=clc]{Descriptive Statistics: Raw and Standardized Variables}}
    \label{tab:descrptstats}
    \begin{threeparttable}
    \begin{tabular}{@{}l S S S S S S S@{}}
        \toprule
        \textbf{Variable} & {\textbf{Mean}} & {\textbf{SD}} & {\textbf{Min}} & {\textbf{Median}} & {\textbf{Max}} & {\textbf{Skew}} & {\textbf{Kurt}} \\
        \midrule 
        \multicolumn{7}{@{}l@{}}{\textit{Panel A: Raw variables (original units)}} \\
        \midrule
        $X_1$ -- GDP/pc       & 62.492  & 17.035  & 26.000 & 63.000  & 96.000  & -0.209 & 2.129  \\
        $X_2$ -- Price Level  & 59.906  & 13.704  & 20.300 & 60.500  & 91.800  & -0.253 & 2.690  \\
        $X_3$ -- FDI          & 77.057  & 66.436  & 0.474  & 65.521  & 393.995 & 1.745  & 7.455  \\
        $X_4$ -- Trade        & 115.017 & 34.865  & 43.555 & 116.296 & 203.929 & 0.113  & 2.117  \\
        $X_5$ -- GCF          & 378.169 & 216.927 & 16.081 & 387.136 & 844.834 & 0.032  & 1.917  \\
        $X_6$ -- Gov. Debt    & 38.370  & 21.711  & 3.900  & 37.400  & 99.000  & 0.492  & 2.511  \\
        $X_7$ -- Priv. Debt   & 69.996  & 28.545  & 9.900  & 68.400  & 146.700 & 0.334  & 2.610  \\
        \midrule
        \multicolumn{7}{@{}l@{}}{\textit{Panel B: Standardized variables (z-scores)}} \\
        \midrule
        $X_1$ -- GDP/pc       &   0.000 &   1.000 &  -2.142 &  0.030  &   1.967 & -0.209 & 2.129 \\
        $X_2$ -- Price Level  &   0.000 &   1.000 &  -2.890 &   0.043 &   2.327 & -0.253 & 2.690 \\
        $X_3$ -- FDI          &   0.000 &   1.000 &  -1.153 &  -0.174 &   4.771 & 1.745  & 7.455 \\
        $X_4$ -- Trade        &   0.000 &   1.000 &  -2.050 &   0.037 &   2.550 & 0.113  & 2.117 \\
        $X_5$ -- GCF          &   0.000 &   1.000 &  -1.700 &   0.041 &   2.151 & 0.032  & 1.917 \\
        $X_6$ -- Gov. Debt    &   0.000 &   1.000 &  -1.588 &  -0.045 &   2.793 & 0.492  & 2.511  \\
        $X_7$ -- Priv. Debt   &   0.000 &   1.000 &  -2.105 &  -0.056 &   2.687 & 0.334  & 2.610 \\
        \bottomrule
    \end{tabular}
    \begin{tablenotes}[flushleft]
        \footnotesize
        \item \textit{Notes:} Panel A reports descriptive statistics for the variables in their original units, as defined in Table~\ref{tab:variables}. Panel B reports the same statistics after standardization, where each variable is transformed as $\tilde{x}_{it} = (x_{it} - \bar{x}) / \sigma_x$, using the pooled sample mean ($\bar{x}$) and standard deviation ($\sigma_x$). Standardized variables have mean zero and unit variance by construction; the minimum and maximum values therefore indicate the range of variation in standard-deviation units.
    \end{tablenotes}
    \end{threeparttable}
\end{table}

{\setlength{\tabcolsep}{6pt}
\begin{table}[!htbp]
\centering
\caption{\added[id=clc]{Pairwise correlations of regressors $X_{1}$--$X_{7}$ (pooled, $n=11 \times 29=319$).}}
\label{tab:corr}
\small
\begin{tabular}{l rrrrrrr}
\toprule
 & $X_{1}$ & $X_{2}$ & $X_{3}$ & $X_{4}$ & $X_{5}$ & $X_{6}$ & $X_{7}$ \\
\midrule
$X_{1}$ & 1.000 &  &  &  &  &  &  \\
$X_{2}$ & 0.801$^{***}$ & 1.000 &  &  &  &  &  \\
$X_{3}$ & 0.349$^{***}$ & 0.338$^{***}$ & 1.000 &  &  &  &  \\
$X_{4}$ & 0.589$^{***}$ & 0.588$^{***}$ & 0.543$^{***}$ & 1.000 &  &  &  \\
$X_{5}$ & 0.692$^{***}$ & 0.698$^{***}$ & 0.690$^{***}$ & 0.638$^{***}$ & 1.000 &  &  \\
$X_{6}$ & 0.220$^{***}$ & 0.164$^{***}$ & 0.192$^{***}$ & 0.176$^{***}$ & 0.274$^{***}$ & 1.000 &  \\
$X_{7}$ & 0.364$^{***}$ & 0.490$^{***}$ & 0.540$^{***}$ & 0.426$^{***}$ & 0.532$^{***}$ & 0.054 & 1.000 \\
\bottomrule
\end{tabular}%
\par\smallskip
\end{table}
}

{\renewcommand{\cellalign}{rc}\setlength{\tabcolsep}{4pt}
\begin{table}[!htbp]
\centering
\caption{\added[id=clc]{Country-level summary statistics, mean (SD in parentheses).}}
\label{tab:countrystats}
\resizebox{\textwidth}{!}{%
\begin{tabular}{l rrrrrrrr}
\toprule
Country & $X_{1}$ & $X_{2}$ & $X_{3}$ & $X_{4}$ & $X_{5}$ & $X_{6}$ & $X_{7}$ & $\Delta y$ \\
\midrule
Bulgaria  & \makecell{43.59 \\ \scriptsize(10.39)} & \makecell{42.73 \\ \scriptsize(10.34)} & \makecell{112.28 \\ \scriptsize(72.87)}  & \makecell{106.15 \\ \scriptsize(22.36)} & \makecell{311.62 \\ \scriptsize(200.14)} & \makecell{37.91 \\ \scriptsize(27.38)} & \makecell{80.52 \\ \scriptsize(42.77)} & \makecell{0.79 \\ \scriptsize(2.37)} \\
Czechia   & \makecell{84.10 \\ \scriptsize(6.46)}  & \makecell{62.78 \\ \scriptsize(12.39)} & \makecell{77.90 \\ \scriptsize(41.69)}   & \makecell{121.19 \\ \scriptsize(24.76)} & \makecell{450.38 \\ \scriptsize(239.79)} & \makecell{30.03 \\ \scriptsize(10.31)} & \makecell{69.57 \\ \scriptsize(11.71)} & \makecell{0.45 \\ \scriptsize(1.94)} \\
Estonia   & \makecell{64.55 \\ \scriptsize(16.69)} & \makecell{67.53 \\ \scriptsize(12.95)} & \makecell{120.28 \\ \scriptsize(72.59)}  & \makecell{143.69 \\ \scriptsize(14.93)} & \makecell{446.48 \\ \scriptsize(249.94)} & \makecell{9.02 \\ \scriptsize(4.70)}   & \makecell{92.47 \\ \scriptsize(30.80)} & \makecell{1.48 \\ \scriptsize(2.79)} \\
Croatia   & \makecell{59.10 \\ \scriptsize(7.81)}  & \makecell{64.48 \\ \scriptsize(3.63)}  & \makecell{58.66 \\ \scriptsize(35.81)}   & \makecell{83.97 \\ \scriptsize(14.00)}  & \makecell{334.41 \\ \scriptsize(193.35)} & \makecell{53.19 \\ \scriptsize(21.49)} & \makecell{82.52 \\ \scriptsize(31.18)} & \makecell{1.10 \\ \scriptsize(1.63)} \\
Latvia    & \makecell{52.21 \\ \scriptsize(13.29)} & \makecell{62.07 \\ \scriptsize(12.17)} & \makecell{65.55 \\ \scriptsize(33.82)}   & \makecell{110.24 \\ \scriptsize(21.72)} & \makecell{404.67 \\ \scriptsize(241.19)} & \makecell{28.25 \\ \scriptsize(15.31)} & \makecell{72.23 \\ \scriptsize(33.08)} & \makecell{1.41 \\ \scriptsize(1.82)} \\
Lithuania & \makecell{61.38 \\ \scriptsize(18.34)} & \makecell{57.10 \\ \scriptsize(11.34)} & \makecell{54.15 \\ \scriptsize(30.37)}   & \makecell{123.34 \\ \scriptsize(27.34)} & \makecell{333.37 \\ \scriptsize(187.27)} & \makecell{28.58 \\ \scriptsize(11.10)} & \makecell{50.21 \\ \scriptsize(17.55)} & \makecell{1.86 \\ \scriptsize(2.37)} \\
Hungary   & \makecell{64.07 \\ \scriptsize(8.05)}  & \makecell{59.20 \\ \scriptsize(6.75)}  & \makecell{168.86 \\ \scriptsize(118.22)} & \makecell{142.65 \\ \scriptsize(27.79)} & \makecell{379.82 \\ \scriptsize(211.08)} & \makecell{69.31 \\ \scriptsize(8.93)}  & \makecell{76.33 \\ \scriptsize(21.99)} & \makecell{0.86 \\ \scriptsize(1.25)} \\
Poland    & \makecell{60.52 \\ \scriptsize(11.69)} & \makecell{57.04 \\ \scriptsize(5.76)}  & \makecell{52.81 \\ \scriptsize(30.20)}   & \makecell{80.37 \\ \scriptsize(21.11)}  & \makecell{327.43 \\ \scriptsize(181.93)} & \makecell{47.48 \\ \scriptsize(5.81)}  & \makecell{56.14 \\ \scriptsize(19.76)} & \makecell{1.21 \\ \scriptsize(1.42)} \\
Romania   & \makecell{48.10 \\ \scriptsize(16.76)} & \makecell{45.64 \\ \scriptsize(9.19)}  & \makecell{52.83 \\ \scriptsize(31.68)}   & \makecell{69.91 \\ \scriptsize(13.50)}  & \makecell{355.71 \\ \scriptsize(215.12)} & \makecell{27.75 \\ \scriptsize(12.54)} & \makecell{49.51 \\ \scriptsize(14.67)} & \makecell{1.66 \\ \scriptsize(2.21)} \\
Slovenia  & \makecell{84.10 \\ \scriptsize(3.74)}  & \makecell{79.48 \\ \scriptsize(4.88)}  & \makecell{28.32 \\ \scriptsize(18.26)}   & \makecell{130.35 \\ \scriptsize(26.18)} & \makecell{390.00 \\ \scriptsize(210.32)} & \makecell{46.12 \\ \scriptsize(24.18)} & \makecell{74.01 \\ \scriptsize(24.94)} & \makecell{0.48 \\ \scriptsize(1.66)} \\
Slovakia  & \makecell{65.69 \\ \scriptsize(10.04)} & \makecell{60.92 \\ \scriptsize(14.19)} & \makecell{55.97 \\ \scriptsize(33.43)}   & \makecell{153.33 \\ \scriptsize(31.58)} & \makecell{425.96 \\ \scriptsize(224.31)} & \makecell{44.43 \\ \scriptsize(10.42)} & \makecell{66.46 \\ \scriptsize(18.38)} & \makecell{0.90 \\ \scriptsize(2.35)} \\
\bottomrule
\end{tabular}%
}
\par\smallskip
\footnotesize\textit{Notes.} Each cell shows the country-specific mean over 1996--2024 with the standard deviation in parentheses below. $T=29$ for every country.
\end{table}
}
 
\section{Results  \label{sec:Results}}

Although contemporary economic research investigators utilize vast amounts of data and sophisticated research methods, this study restricted itself to seven macroeconomic factors of economic growth, as described in the previous section, and to eleven countries from Central and Eastern Europe. Such conscious restriction may be helpful for the reasonable interpretation of results. Indeed, all countries under investigation undergo similar political and economic transformations as they join the European Economic Area and share many common development features. Thus, the linear regression with varying coefficients, as described in Section \ref{Eq:regression}, may be productive in comparing the contributions to economic growth of the seven selected macroeconomic and financial factors.  

\added[id=clc]{We do acknowledge the problem of statistical accuracy, implementing this analysis with such a small number of countries in a limited number of years. The statistics of error terms (residuals) in Eqs. \eqref{Eq:regression} and \eqref{eq:LRVC_P}, $e_{i,t}$, is a primary source of information on the method's accuracy. In Table \ref{table:statresiduals} we present basic descriptive statistics of $e_{i,t}$, calculated from the total number of values $11 \times 29 = 319$, and separately for countries, based on 29 error values per country. The descriptive statistics reported in the first row of Table \ref{table:statresiduals} indicate an acceptable overall fit of the proposed model The mean of all residuals is equal to zero with high precision, and the median deviates from zero only by \SI{1.7}{\percent} of SD. The observed skewness is moderate with more considerable deviation of kurtosis $Kurt=4.549$ from the value $3.$ of the Normal distribution. The symmetry is confirmed by $|Q25| \simeq Q75$ as well. The $IQR/SD$ ratio of $0.98$, which is below $1.35$, indicates deviations in the tails relative to a Normal distribution. This deviation from normality is confirmed by $RMSE/MAE \simeq 1.402$ deviating from $1.25$. The Jarque-Bera test of normality gives a p-value of $0.0172$, indicating a statistically significant departure from normality at the 5\% level.} 

\added[id=clc]{Limited data basis for accurate statistics appears more considerable when we investigate country data separately. For the two countries, Czechia and Latvia, the residuals' means deviate considerably from zero. These deviations are in opposite directions, annihilating in a wider view, but have a direct impact on $\Delta y$ decomposition into the contribution of factors under investigation. In the proposed method for evaluating the contribution of selected factors, the deviation of the residual mean from zero has a significant impact. Nevertheless, this affects only the results of two countries. Other statistical indicators fluctuate around the values characterizing the Normal distribution. Though fluctuations are considerable, they cause smaller deviations of the model $y$ from the empirical GDP time series.}

\begin{table}[!htbp]
    \centering
    \footnotesize                                
    \setlength{\tabcolsep}{3pt}                  
    \sisetup{
        table-format        = -1.3,              
        table-number-alignment = center,
        detect-weight       = true,
        detect-family       = true
    }
    \caption{\added[id=clc]{Descriptive Statistics of Residuals $e_{i,t}$ in Regression \eqref{eq:LRVC_P}}}
    \label{table:statresiduals}
 
    \begin{threeparttable}
    \begin{tabular*}{\textwidth}{@{\extracolsep{\fill}} l S S S S S S S S S @{}}
        \toprule
        \textbf{Variable} & {\textbf{Mean}} & {\textbf{SD}} & {\textbf{Median}} & {\textbf{Skew}} & {\textbf{Kurt}} & {\textbf{Q25}} & {\textbf{Q75}} & {\textbf{IQR}} & {\textbf{RMSE/MAE}} \\
        \midrule
        \multicolumn{10}{@{}l}{\textit{Panel A: Pooled residuals (319 observations)}} \\
        \midrule
        $e_{i,t}$ -- all  &  0.000 &  0.747 &  0.013 &  0.037 & 4.549 & -0.359 & 0.372 & 0.731 & 1.402 \\
        \midrule
        \multicolumn{10}{@{}l}{\textit{Panel B: Residuals by country (29 observations each)}} \\
        \midrule
        Bulgaria   &  0.062 & 0.643 &  0.011 &  0.263 & 3.233 & -0.267 & 0.410 & 0.678 & 1.421 \\
        Czechia    &  0.192 & 0.492 &  0.145 &  1.156 & 4.835 & -0.096 & 0.320 & 0.416 & 1.388 \\
        Estonia    & -0.027 & 0.543 & -0.015 & -0.399 & 2.411 & -0.370 & 0.450 & 0.820 & 1.239 \\
        Croatia    &  0.025 & 0.820 &  0.171 & -0.661 & 3.862 & -0.157 & 0.224 & 0.381 & 1.407 \\
        Latvia     & -0.207 & 0.654 & -0.124 & -0.289 & 2.716 & -0.743 & 0.354 & 1.098 & 1.245 \\
        Lithuania  &  0.053 & 0.680 &  0.187 & -0.798 & 3.594 & -0.167 & 0.385 & 0.553 & 1.324 \\
        Hungary    & -0.066 & 0.407 & -0.024 & -0.608 & 3.000 & -0.229 & 0.121 & 0.350 & 1.366 \\
        Poland     & -0.014 & 1.275 & -0.111 &  0.407 & 2.850 & -0.729 & 0.491 & 1.221 & 1.332 \\
        Romania    & -0.005 & 0.793 & -0.031 &  0.431 & 3.231 & -0.506 & 0.464 & 0.970 & 1.291 \\
        Slovenia   &  0.037 & 0.817 & -0.015 &  0.045 & 2.308 & -0.417 & 0.768 & 1.186 & 1.257 \\
        Slovakia   & -0.049 & 0.799 & -0.006 & -0.442 & 5.649 & -0.320 & 0.211 & 0.531 & 1.581 \\
        \bottomrule
    \end{tabular*}
 
    \begin{tablenotes}[flushleft]
        \footnotesize
        \item \textit{Notes:} Panel A reports descriptive statistics for the residuals computed from all 319 observations. Panel B reports the same statistics for each country separately, computed from 29 observations. Four indicators are added relative to Table~\ref{tab:descrptstats}: Q25 is the first quartile (the value below which \SI{25}{\percent} of the residuals fall); Q75 is the third quartile (the value below which \SI{75}{\percent} of the residuals fall); IQR is the interquartile range (Q75~$-$~Q25); RMSE/MAE is the ratio of the root mean squared error to the mean absolute error.
    \end{tablenotes}
    \end{threeparttable}
\end{table}

\begin{figure}[h]
\centering
\includegraphics[width=0.5\textwidth]{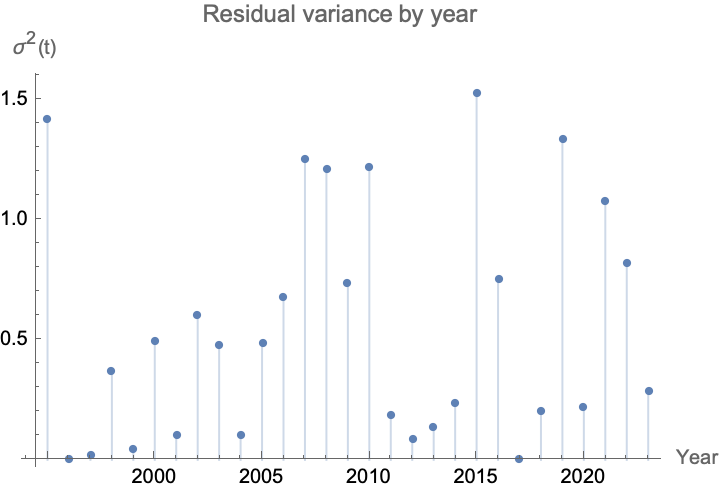}
\caption{Residual variance $\sigma^2(t)=\frac{1}{N} \sum_{i=1}^N e_{i,t}^2$ calculated in the linear regression Eq.\ref{eq:LRVC_P} 
\label{fig3-2}}
\end{figure}

\added[id=clc]{Bartlett's test of heteroskedasticity rejects the null of equal residual variances across the 29 yearly cross-sections (B = 193.09, df = 28, p < 0.001), confirming substantial time-wise heteroskedasticity. The residual variance spikes around the 2008–09 financial crisis and the 2020 COVID shock (Figure \ref{fig3-2}). The ratio of largest to smallest country variance is $(1.275/0.407)^2 \simeq 9.8$, see Table \ref{table:statresiduals},
which is large and almost certainly rejects the equal-variance null.}

\added[id=clc]{Common for all countries, time-varying coefficients $\alpha_t$ and $\beta_{t,j}$ provide the most valuable information about the dynamics of CEE development. We have to acknowledge that the lack of data for the comprehensive statistical analysis is the main obstacle in our research. The method for dealing with moderate multicollinearity of regressors incorporates the variable transformations described in subsection \ref{sec:orthog} and the back-transformation that reverts to the original regressors. The transformation slightly complicates the statistical analysis of the evaluated coefficients $\alpha_t$, and $\beta_{t,j}$. Results of the coefficients and the estimated standard errors for the selected number of years are given in the Table \ref{tab:tvcoefs}. Despite considerable uncertainty in the defined coefficients, the modeled dynamics of $y(t)$ reproduce the empirical time series of GDP for almost all countries investigated. Only for Czechia and Latvia, accumulated errors are higher; see Table \ref{table1}, column $\bm{e_{i,t=2024}^a}$.}

\begin{figure}[h]
\centering
\includegraphics[width=0.3\textwidth]{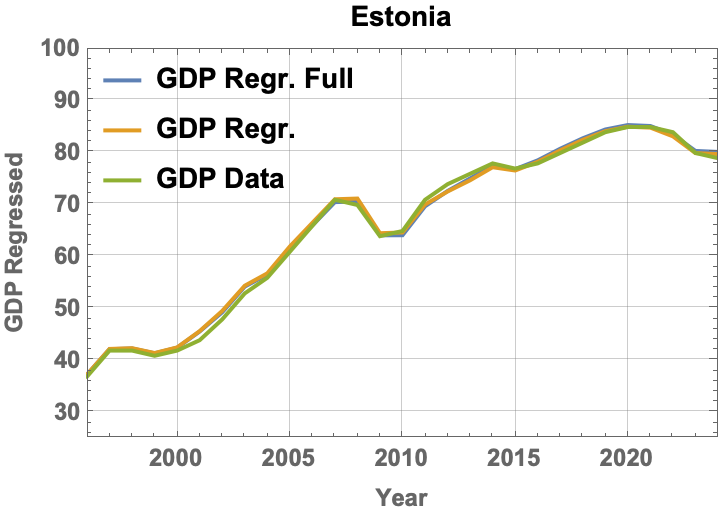}
\includegraphics[width=0.3\textwidth]{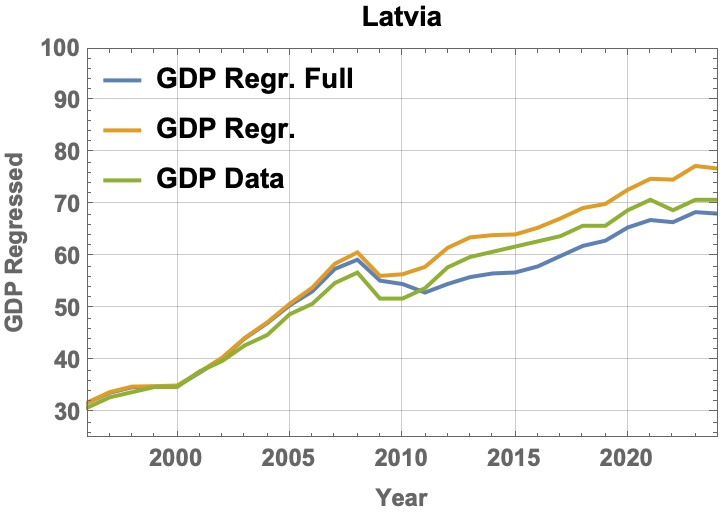}
\includegraphics[width=0.3\textwidth]{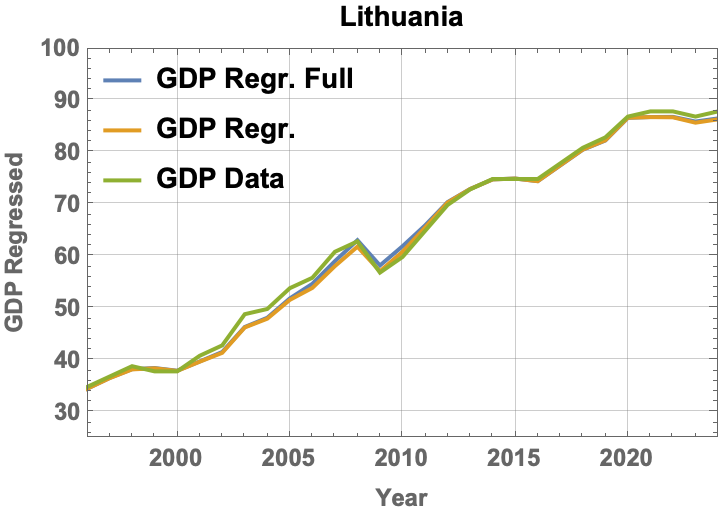}
\caption{The comparison of regressed GDP's for the Baltic countries with statistical data.
\label{fig4}}
\end{figure}

Seeking to evaluate the accuracy of the linear regression with time-varying coefficients method for our investigation, we compare three time series as defined in equations \eqref{Eq:GDPData}, \eqref{Eq:GDPRegr}, \eqref{Eq:GDPRefrFull}, see three sub-figures for the Baltic countries in Fig. \ref{fig4}. Errors accumulate slightly in the case of Latvia, but regression for Estonia and Lithuania yields a very accurate replication of the empirical economic growth series. For the other CEE countries, the time series behave similarly; we will provide more information in the following table.

\begin{table}[!htbp]
    \centering
    \footnotesize
    \setlength{\tabcolsep}{4pt}
    \caption{\added[id=clc]{Time-Varying Coefficients $\beta_{t,j}$ and $\alpha_t$ for Selected Years}}
    \label{tab:tvcoefs}

    \begin{threeparttable}
    \begin{tabular*}{\textwidth}{@{\extracolsep{\fill}} l c c c c c c c @{}}
        \toprule
        \textbf{Regressor} & \textbf{1997} & \textbf{2002} & \textbf{2009} & \textbf{2013} & \textbf{2018} & \textbf{2021} & \textbf{2024} \\
        \midrule
        $\alpha_t$
            & $-5.799^{***}$ & $\phantom{-}0.300$ & $\phantom{-}12.081$ & $\phantom{-}3.788$ & $\phantom{-}7.586^{***}$ & $\phantom{-}16.146^{**}$ & $\phantom{-}4.713$  \\
            & $(0.407)$ & $(2.354)$ & $(5.949)$ & $(2.053)$ & $(0.477)$ & $(3.362)$ & $(4.429)$ \\
        \addlinespace
$X_1$ & $-0.128^{***}$ & $-0.105^{**}$ & $\phantom{-}0.125$ & $-0.186^{**}$ & $-0.013^{*}$ & $-0.168^{**}$ & $\phantom{-}0.034$ \\
            & $(0.007)$ & $(0.026)$ & $(0.108)$ & $(0.041)$ & $(0.006)$ & $(0.035)$ & $(0.071)$ \\
        \addlinespace
        $X_2$ & $\phantom{-}0.149^{***}$ & $\phantom{-}0.036$ & $-0.526^{*}$ & $\phantom{-}0.512^{***}$ & $\phantom{-}0.123^{***}$ & $\phantom{-}0.223^{**}$ & $-0.104$ \\
            & $(0.008)$ & $(0.042)$ & $(0.213)$ & $(0.082$ & $(0.010)$ & $(0.070)$ & $(0.094)$ \\
        \addlinespace
        $X_3$ & $-0.023$ & $-0.024$ & $-0.032$ & $\phantom{-}0.058^{**}$ & $\phantom{-}0.014^{***}$ & $\phantom{-}0.006$ & $-0.004$ \\
            & $(0.014)$ & $(0.021)$ & $(0.038)$ & $(0.010)$ & $(0.001)$ & $(0.005)$ & $(0.005)$ \\
        \addlinespace
        $X_4$ & $\phantom{-}0.021^{***}$ & $\phantom{-}0.025$ & $-0.062$ & $-0.015$ & $-0.022^{***}$ & $-0.035$ & $\phantom{-}0.014$ \\
            & $(0.002)$ & $(0.013)$ & $(0.034)$ & $(0.010)$ & $(0.002)$ & $(0.019)$ & $(0.023)$ \\
        \addlinespace
        $X_5$ & $\phantom{-}0.166^{***}$ & $-0.002$ & $\phantom{-}0.044$ & $-0.029^{**}$ & $-0.009^{***}$ & $-0.020^{**}$ & $-0.003$ \\
            & $(0.010)$ & $(0.012)$ & $(0.031)$ & $(0.006)$ & $(0.001)$ & $(0.006)$ & $(0.006)$ \\
        \addlinespace
        $X_6$ & $-0.002$ & $\phantom{-}0.014$ & $\phantom{-}0.225^{*}$ & $-0.042^{**}$ & $-0.021^{***}$ & $-0.009$ & $-0.021$ \\
            & $(0.004)$ & $(0.016)$ & $(0.086)$ & $(0.013)$ & $(0.003)$ & $(0.015)$ & $(0.024)$ \\
        \addlinespace
        $X_7$ & $-0.082^{***}$ & $\phantom{-}0.069$ & $\phantom{-}0.009$ & $-0.125^{**}$ & $-0.068^{***}$ & $-0.012$ & $\phantom{-}0.039$ \\
            & $(0.006)$ & $(0.040)$ & $(0.042)$ & $(0.021)$ & $(0.003)$ & $(0.021)$ & $(0.033)$ \\
        \addlinespace
        \midrule
        $R^2$              & $0.999$ & $0.935$ & $0.886$ & $0.959$ & $0.995$ & $0.943$ & $0.642$ \\
        Adjusted $R^2$     & $0.997$ & $0.783$ & $0.620$ & $0.863$ & $0.983$ & $0.810$ & $-0.193$ \\
        Residual SD        & $0.162$ & $0.606$ & $2.109$ & $0.555$ & $0.124$ & $0.898$ & $1.021$ \\
        F-statistic        & $428.1$ & $6.165$ & $3.331$ & $10.02$ & $85.29$ & $7.090$ & $0.769$ \\
        \bottomrule
    \end{tabular*}
    \begin{tablenotes}[flushleft]
        \footnotesize
        \item \textit{Notes:} Each column reports the coefficients of the time-varying cross-sectional regression for the year indicated, recovered in the original $X$-space from the principal-components regression specified in Equation~\eqref{eq:LRVC_P}. Standard errors in parentheses are obtained by the delta method applied to the P-space covariance matrix through the linear back-transformation. Significance levels: $^{*}\,p<0.10$, $^{**}\,p<0.05$, $^{***}\,p<0.01$. $R^2$, Adjusted $R^2$ as $1 - (1-R^2) (n-1)/(n-d-1)$, residual standard deviation, and the F-statistic refer to the yearly fit in the original principal-components space.
    \end{tablenotes}
    \end{threeparttable}
\end{table}

\begin{figure}[h]
\centering
\includegraphics[width=0.3\textwidth]{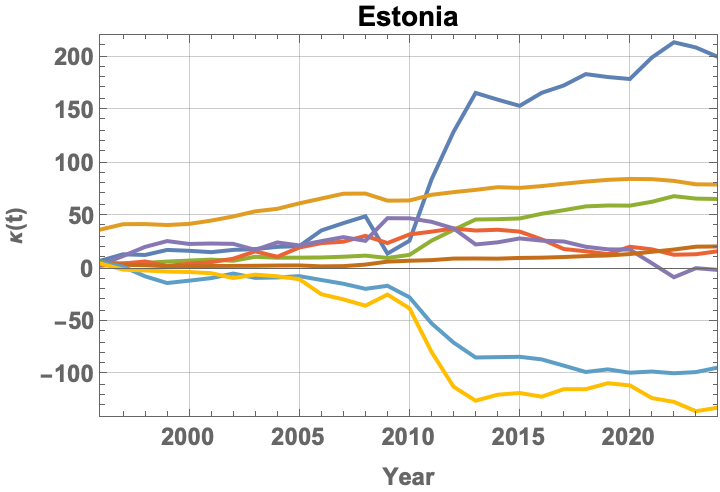}
\includegraphics[width=0.3\textwidth]{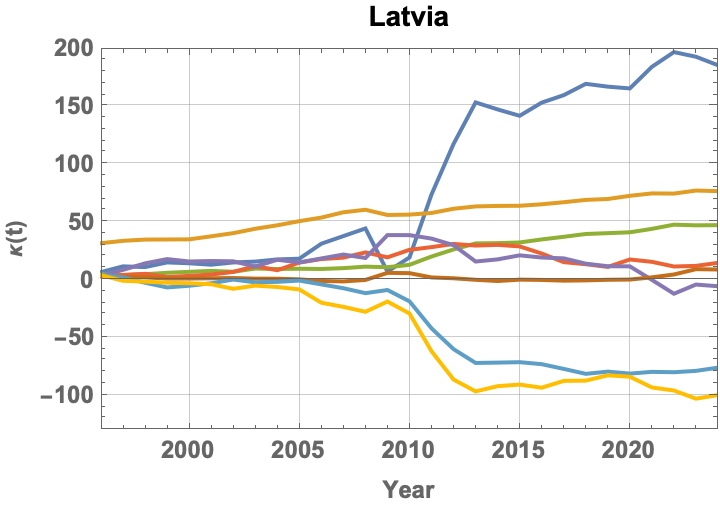}
\includegraphics[width=0.35\textwidth]{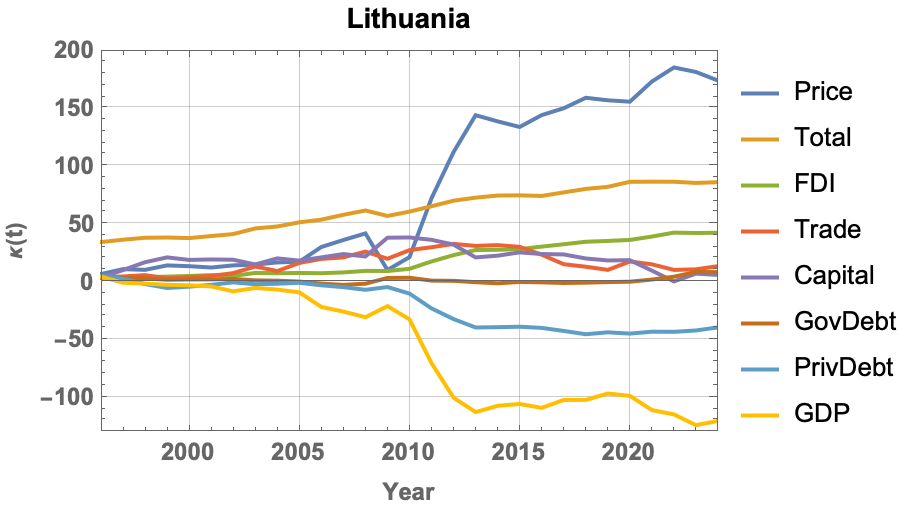}
\caption{The comparison of component contributions to the GDPs of the Baltic countries.
\label{fig5}}
\end{figure}

The primary purpose of this research is to evaluate the contribution of seven factors to economic growth. We calculate component contributions using Eq. \eqref{Eq:components2}; see the results for the Baltic countries in Fig. \ref{fig5}. From a macroeconomic perspective, prices are the primary driver of economic growth. The role of prices is consistent with the Penn effect. The primary opposing force to economic growth is the level of economic development achieved, consistent with neoclassical and other economic growth models. The contribution of other factors is considerably lower; nevertheless, foreign direct investment is one of the main forces of growth for all the countries considered. The positive contribution of trade is considerable as well, but can be comparable to capital formation. The role of Centr. Government debt fluctuates between negative and positive values and is probably more closely related to the government's economic policy.     
The primary outcome of this consideration is a sufficiently high negative contribution of private debt, as one can see from Fig.  \ref{fig5} and the Table \ref{table1}. The different private debt developments in the Baltic countries, as shown in the second sub-figure of Fig. \ref{fig3}, may serve as one possible explanation for the differences in economic development. Private debt's negative impact on the development of the Baltic countries is directly related to the level of private debt achieved, as shown in Fig. \ref{fig5}. The differences in the contributions of other factors are considerably lower; thus, our research can serve as evidence of a much more negative effect of private debt on the development of Estonia and Latvia compared with Lithuania. The particular role of private debt is also evident in other CEE countries, as shown in Table \ref{table1}. 

\begin{table}[!htbp] 
\centering
\caption{The component $j$ contributions to the GDP in year 2024, $\kappa_{i,2024,j}$, for the all CEE countries considered. The column order is determined by the average contribution $Xi$. Two additional columns: the contributions' sum, \textbf{Total}, and accumulated error term $\bm{e_{i,t=2024}^a}$, see Eq. \eqref{Eq:errorAccumulated}, are included.} \label{table1}
\begingroup
\scriptsize                
\setlength{\tabcolsep}{3pt}
\renewcommand{\arraystretch}{0.9}

\begin{tabularx}{\textwidth}{m{0.9cm}>{\centering\arraybackslash}X>{\centering\arraybackslash}X>{\centering\arraybackslash}X>{\centering\arraybackslash}X>{\centering\arraybackslash}X>{\centering\arraybackslash}X>{\centering\arraybackslash}X>{\centering\arraybackslash}X>{\centering\arraybackslash}X}
\midrule
\multicolumn{1}{X}{{\textbf{Country}}} & \multicolumn{1}{X}{{\textbf{  X2}}} & \multicolumn{1}{X}{{~~\textbf{Total}}} & \multicolumn{1}{X}{{\textbf{X3}}} & \multicolumn{1}{X}{{\textbf{X4}}} & \multicolumn{1}{X}{{~\textbf{X5}}} & \multicolumn{1}{X}{{~\textbf{X6}}} & \multicolumn{1}{X}{{\textbf{X7}}} & \multicolumn{1}{X}{{\textbf{X1}}} & \multicolumn{1}{X}{{~~$\bm{e_{i,t=2024}^a}$}}\\ \midrule
\text{Bulgaria} &   $143.0$ & $64.2$ & $64.3$ & $16.6$ & $-2.5$ & $5.8$ & $-84.7$ & $-78.2$ & $1.8$  \\  
\text{Czechia} &   $206.2$ & $85.4$ & $57.3$ & $20.2$ & $9.7$ & $14.6$ & $-52.7$ & $-169.9$ & $5.6$    \\ 
\text{Estonia}    &   $200.2$ & $79.8$ & $66.1$ & $16.8$ & $-0.9$ & $21.3$ & $-93.0$ & $-130.8$ & $-0.8$   \\ 
\text{Croatia}   &   $202.7$ & $76.3$ & $45.2$ & $21.4$ & $4.9$ & $-1.4$ & $-80.0$ & $-116.6$ & $0.7$  \\ 
\text{Latvia}   &   $186.0$ & $77.0$ & $47.6$ & $14.9$ & $-5.3$ & $9.2$ & $-75.8$ & $-99.5$ & $-6.0$  \\ 
\text{Lithuania}   &   $174.4$ & $86.5$ & $42.7$ & $13.8$ & $6.2$ & $8.7$ & $-39.2$ & $-120.2$ & $1.5$ \\ 
\text{Hungary}  &   $181.6$ & $78.9$ & $81.5$ & $16.5$ & $5.9$ & $-9.5$ & $-70.5$ & $-126.5$ & $-1.9$ \\ 
\text{Poland}  &   $173.8$ & $79.4$ & $44.0$ & $17.5$ & $7.1$ & $-0.5$ & $-41.3$ & $-121.2$ & $-0.4$  \\
\text{Romania}  &   $144.5$ & $78.2$ & $41.8$ & $17.1$ & $1.2$ & $10.8$ & $-42.7$ & $-94.5$ & $-0.2$  \\
\text{Slovenia}  &   $248.0$ & $89.9$ & $39.3$ & $19.5$ & $9.3$ & $8.8$ & $-66.2$ & $-168.7$ & $1.1$  \\
\text{Slovakia}  &   $195.1$ & $76.4$ & $45.2$ & $13.5$ & $5.9$ & $4.3$ & $-48.3$ & $-139.3$ & $-1.4$  \\  \bottomrule
\end{tabularx}
\endgroup
\end{table}  

\begin{figure}[h]
\centering
\includegraphics[width=0.75\textwidth]{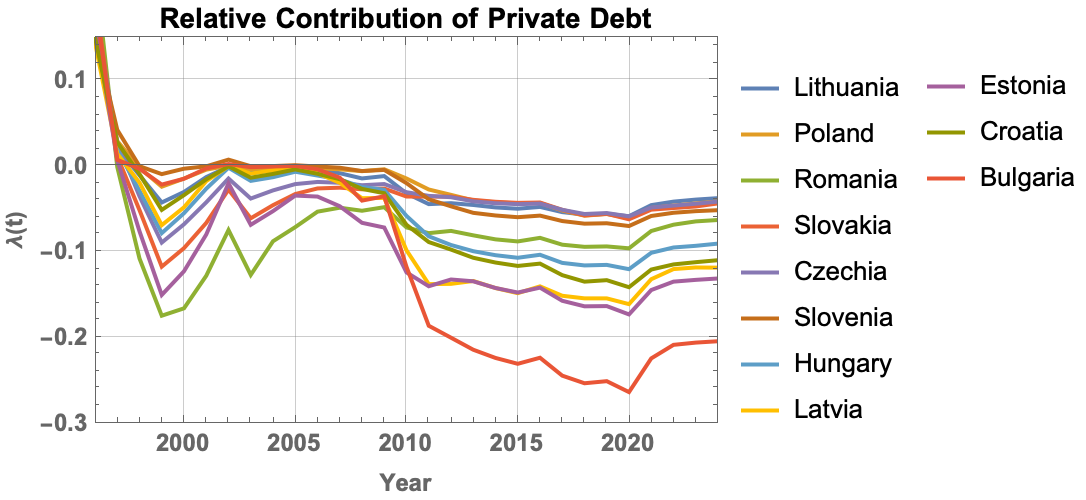}
\caption{The relative contribution of private debt to the development of CEE countries $\lambda(t)$ as defined in Eq. \ref{Eq:relcontr7}. The order of the legends in this Figure is descending according to the values given in column \textbf{X7} of the Table \ref{table1}. 
\label{fig6}}
\end{figure}

We examine the specific impact of private debt on the economic growth of CEE countries, particularly in the post-crisis period, as defined in Eq. \eqref{Eq:relcontr7}, and illustrated in Fig. \ref{fig6},  which plots the relative contribution of private debt.
From our perspective, the private debt contribution presented in Fig. \ref{fig6} and Table \ref{table1} explains many peculiarities of CEE economic development dynamics, as reflected in the empirical data and visualized in Fig. \ref{fig2} by the rank change of countries. 
\section{Discussion and conclusions  \label{sec:Discussion}}
\subsection{\added[id=clc]{Main results \label{sec:mainres }}}
We investigated the dynamics of economic growth in Central and Eastern Europe from the perspective of long-range macroeconomic processes. In this study, we employ the widely accepted Purchasing Power Parity methodology for international comparison. We use data from recognized, publicly available sources, allowing our results to be easily replicated. This panel research is restricted to $N=11$ countries that have acceded to the European Union and have very similar external development conditions. A smaller panel of Baltic countries shares the same heritage of political transformations and serves as a more specific interest of ours. The conscious decision to restrict the number of countries affects our choice to limit the number of macroeconomic regressors to $d=7$ for this panel investigation, as $d<N$ is a method requirement.

Many previous studies and our results confirm the rapid convergence of CEE countries with Western Europe, which is historically more advanced. However, the pace of this convergence varies considerably among the countries investigated. We aim to identify the most reliable macroeconomic factors that explain differences in the speed of economic growth and convergence. The Baltic countries, which had followed very similar development trajectories before the 2008 global financial crisis, later significantly changed course. The impact of World crises is quantitatively evident in our investigation, confirming the necessity to use regression with varying coefficients. The development of Lithuania, as projected by the IMF WEO data, has sparked our interest in finding a reasonable explanation for this success. 

The regression with varying coefficients described in Section \ref{sec:regression} provides us the decomposition of $\Delta y_{i,t}$ time series into the sub-series $\Delta y_{i,t,j}$ of $d=7$ components: Achieved level of GDP, Price level, FDI, Trade, Gross capital formation, Central government debt, Private debt. The decompositions for all countries under investigation are qualitatively similar; see the example of the Baltic countries in Fig. \ref{fig5}. Results confirm previous findings that FDI, international trade, and capital formation are the main drivers of development. The contribution of governmental debt fluctuates quantitatively around zero and appears most positive for Estonia, which has the lowest government debt level. \added[id=clc]{Thus, the quantitative component contributions should be interpreted as relative descriptive estimates, rather than causal effects, and compared across the panel countries.} The main quantitative contributions to GDP growth come from the positive effects of the achieved price level and the adverse effects of the achieved GDP level. It is in agreement with the Penn effect \citep{Hassan2016JIE} and the classical Solow growth model \citep{Ertur2007JAE}. 
\subsection{\added[id=clc]{Policy implications} \label{sec:policImpl}}
The contribution of private debt appears negative across all countries, ranging from the highest in Lithuania, Poland, and Romania to the lowest in Estonia. The fluctuations in private debt contributions are the most considerable, varying by more than twice. Thus, we consider private debt to be the main determining factor of economic growth in CEE countries. To quantify and visualize the contribution of private debt, we introduced the parameter $\gamma_i(t)$ in Eq. \ref{Eq:relcontr7} and plotted the results in Fig. \ref{fig6}. The importance of private debt grew significantly after the 2008 global financial crisis and has become a key factor in economic success in CEE. \added[id=clc]{It should be evident that crises considerably increased the price of the debt, reducing investment and growth opportunities. We consider this result the main finding of our research, with potential policy implications. Lithuania's success trajectory suggests that keeping private debt growth in line with GDP growth is a replicable policy. We will continue our investigation of economic growth data by applying more advanced methods and extending the scope to include additional countries and regions. From our perspective, it is valuable to consider the role of private debt in the broader theories of economic growth. The agent-based modeling of economic convergence \cite{Gontis2026Arxiv} potentially has room to incorporate financial components. Empirical analysis of economic growth data combining macroeconomic modeling with econometric methods should lead to more targeted policy implications.} 
\subsection{\added[id=clc]{Limitations} \label{sec:limitations}}
\added[id=clc]{Despite the limited number of investigated countries, regressors, and years, our results appear reasonable, as the regressed GDPs (see Fig. \ref{fig4} and Table \ref{table1}) reproduce empirical data with acceptable precision. Nevertheless, we acknowledge the limitations of using some contemporary statistical methods, given the very limited data available for the LRVC. The task defines the very low degree of freedom $N-d-1=3$, and we are not able to overcome the challenge. First, other EU countries have different development trajectories, so extending the country list does not increase precision. Second, a smaller number of regressors decreases precision. The selected choice appears optimal, as our experimentation outside the content of this presentation showed.} 

\added[id=clc]{Furthermore, it is evident that the regressors (predictors) in this study are not independent; for example, capital formation is likely to be influenced by credit availability and private debt. The macroeconomic relationship between price levels and GDP is well defined by Penn's law. Our numerical experiments, excluding one obviously related regressor, gave worse reproduction of the empirical GDP time series. Thus, we used the procedure of variable orthogonalization described in subsection \ref{sec:orthog} and reverted to the original variables after implementing OLS calculations. Despite limitations that reduce statistical accuracy, the method provides additional insight into the dynamics of GDP. Results do not contradict the findings of other authors presented in the literature review. A more extensive empirical analysis of World economic growth data would help develop the method. Nevertheless, our study yields new insights into the diversity of economic growth across CEE countries, warranting further investigation and discussion.}
  
\section{Abbreviations}
The following abbreviations are used in this manuscript:\\
 
\noindent 
\begin{tabular}{@{}ll}
CEE   & Central and Eastern Europe\\
ESTAT & Eurostat\\
EU    & European Union\\
FDI   & Foreign direct investment\\
IMF   & International Monetary Fund\\
GCF   & Gross capital formation\\
GDP   & Gross domestic product\\
LRVC  & Linear regression with time-varying coefficients\\
MAE   & Mean absolute error  \\
MPD   & Maddison Project Database\\
PPP   & Purchasing power parity\\
PPS   & Purchasing power standards\\
PWT   & Penn World Tables\\
RMSE  & Root mean squared error \\
TFA   & Technological Frontier Area\\
WB    & World Bank\\
WEO   & World Economic Outlook
\end{tabular}


\section*{About the authors}

\noindent\textbf{Vygintas Gontis} is an Affiliated Professor at the Institute of Theoretical Physics and Astronomy, Vilnius University. He received his Ph.D. in Theoretical and Mathematical Physics from Vilnius University in 1985 and was awarded the Habilitation in 2007 for his thesis on quantum--classical correspondence in dynamically chaotic systems, fractal point processes, and their applications. His current research lies at the interface of statistical physics and finance, with a focus on agent-based and nonlinear stochastic modeling of financial markets, long-range memory, and the bursty behavior of returns and trading activity. In 1991--1993, Vygintas Gontis was working at the Government of Lithuania as Director-General of the Agency for Science, Higher Education, and Technologies. In 2014, he held an eleven-month research internship at the Center for Polymer Studies, Boston University, supported by the Baltic American Freedom Foundation. He has served as President of the Lithuanian Scientific Society and is an Academic Editor for PLOS ONE.

\noindent\textbf{Lesya Kolinets} is an Associate Professor and Senior Research Fellow at the Department of Economics Engineering, Vilnius Gediminas Technical University, Lithuania, where she has worked since 2023. She holds a Ph.D. (2005) and a Doctor of Science degree (Habilitation, 2018) in World Economy and International Economic Relations. From 2019 to 2023, she was a Professor at the Department of International Economics at West Ukrainian National University. Her research interests include international economics, global economic and financial transformations, and sustainable development. She has participated in research projects focused on regional economic integration, global economic development, and the evolving world economic order. Lesya Kolinets is an Academician of the Academy of Economic Sciences of Ukraine and a member of the Lithuanian Scientific Society.

\end{document}